\documentclass[aps,pre,reprint,superscriptaddress,longbibliography,floatfix,showkeys]{revtex4-2}
\usepackage{graphicx}
\usepackage{multirow}
\usepackage{bm}
\usepackage{amsmath}
\usepackage{amssymb}
\usepackage[caption=false]{subfig}
\usepackage{booktabs}


\begin{document}

\title{Sequential Autoregressive Block Metropolis--Hastings for
Two-Dimensional $\phi^4$ Lattice Fields}

\author{Moxian Qian}
\email{qianmoxian@cuhk.edu.cn}
\affiliation{School of Science and Engineering, The Chinese University of Hong Kong, Shenzhen, Guangdong 518172, China}

\author{Shiyang Chen}
\email{2222895@swansea.ac.uk}
\affiliation{Centre for Quantum Fields and Gravity, Department of Physics, Swansea University, SA2 8PP, Swansea, United Kingdom}

\begin{abstract}
We construct a Metropolis-corrected sequential autoregressive block update
for continuous lattice fields. A variational autoregressive network (VAN)
regenerates the sites of an ordered geometric block, and the Hastings ratio
uses forward generation and reverse restoration probabilities evaluated
along their respective partial configurations. We apply the update to the
two-dimensional $\phi^4$ theory at $L=6$, $8$, and $10$, comparing learned
single-site, finite-block, and global proposals. For each trained VAN, the
autocorrelation of $|M|$, measured per sweep, varies nonmonotonically with
block size: intermediate blocks yield smaller point estimates of
$\tau_{\rm int}(|M|)$ than the learned single-site update, while larger
finite blocks have lower acceptance. When networks trained at
$\kappa=0.270$ are applied without retraining to a target at
$\kappa=0.273$, finite-block acceptances change little, whereas the
global-proposal acceptance decreases and its estimated
$\tau_{\rm int}(|M|)$ increases, with the strongest degradation at
$L=10$.
\end{abstract}


\maketitle
\raggedbottom

\section{Introduction}

Markov-chain Monte Carlo simulations can develop long autocorrelation times near continuous phase transitions~\cite{Schaefer2011,Wolff1990}. Nonlocal cluster algorithms provide effective alternatives when suitable model-specific constructions are available~\cite{SwendsenWang1987,Wolff1989,BrowerTamayo1989}. Here we investigate learned proposals whose spatial scope ranges from a single lattice site to the full field, with the resulting transitions corrected by an exact Metropolis--Hastings step~\cite{Metropolis1953,Hastings1970}.

Learned generative models have been used to construct global proposals for lattice field theories, including normalizing-flow samplers combined with importance reweighting or Metropolis correction \cite{Albergo2019,Nicoli2020,Nicoli2021,Abbott2022,Kanwar2020}. Because a global proposal replaces the complete field in one step, its acceptance depends on the accuracy of the learned density over the full configuration. We therefore vary the number of lattice sites regenerated in one Metropolis--Hastings proposal. Autoregressive density estimators factorize a joint density into tractable sitewise conditionals~\cite{vandenoord2016pixel,germain2015made,uria2013rnade}. Masked autoregressive transformations extend this construction within normalizing flows~\cite{papamakarios2017maf}. Autoregressive factorizations have been used as variational samplers and Metropolis proposals for lattice spin systems~\cite{wu2019solving,Wang2020cman,Bialas2022han,Bialas2023potts}. Wu, Rossi, and Carleo construct Metropolis-corrected cluster updates by resampling a variable-length suffix of the autoregressive order and combining the update with symmetry transformations~\cite{Wu2021cluster}. Domain-decomposed flow methods restrict invertible flow proposals to subdomains and apply correction steps at larger scales \cite{Finkenrath2022,Finkenrath2024}; autoregressive sequences of conditional normalizing flows provide a related continuous-field construction~\cite{Dinesh2023lacnf}. A periodic geometric block need not be a suffix of the global autoregressive order: some variables that occur later in that order remain unchanged. Its forward generation and reverse restoration probabilities must therefore be evaluated on different sequences of partial configurations. We derive the corresponding Metropolis--Hastings transition, and Appendix~\ref{app:block_mh_proof} proves detailed balance for each prescribed block and for sweeps whose anchors are sampled independently of the field.

We test the construction in the two-dimensional $\phi^4$ theory using one VAN per volume trained at $\kappa=0.270$. To probe how proposal scope responds to a controlled target-coupling shift, we apply the same networks without retraining at $\kappa_{\rm target}=0.273$, where they supply the single-site, finite-block, and global proposals compared at $L=6$, $8$, and $10$. We also scan $L=6$ with $\kappa_{\rm train}=\kappa_{\rm target}$ and evaluate two acceptance rules on common $L=8$ proposals. Hybrid Monte Carlo (HMC) supplies independent equilibrium reference values~\cite{Duane1987}.

The remainder of this paper is organized as follows. Section~\ref{sec:variational_autoregressive_network} introduces the variational autoregressive network and the learned single-site Metropolis--Hastings update. Section~\ref{sec:block_update} develops the sequential autoregressive block proposal and derives the corresponding forward and reverse proposal probabilities entering the Metropolis--Hastings ratio. Section~\ref{sec:phi4} presents the application to the two-dimensional $\phi^4$ theory, including the numerical setup, physical observables, acceptance rates, autocorrelation times, importance-sampling effective sample size, and the dependence on block size. Section~\ref{sec:conclusion} summarizes the main results and discusses the implications for local, finite-block, and global autoregressive proposals. Additional details on the network architecture, detailed balance, target-coupling dependence, autocorrelation functions, and numerical results are provided in Appendices~\ref{app:architecture}--\ref{app:complete_observables}.

\section{Variational Autoregressive Network}
\label{sec:variational_autoregressive_network}

A variational autoregressive network factorizes the joint density into normalized sitewise conditionals,
\begin{equation}
    p_{\theta}(\boldsymbol{s})
    =
    \prod_{i=1}^{V}
    p_{\theta}(s_i\mid s_{<i}),
    \label{eq:van_factorization}
\end{equation}
where $\boldsymbol{s}=(s_1,\ldots,s_V)$ denotes a configuration, $s_{<i}=\{s_1,\ldots,s_{i-1}\}$, and $\theta$ denotes the network parameters. The log density is
\begin{equation}
    \log p_{\theta}(\boldsymbol{s})
    =
    \sum_{i=1}^{V}
    \log p_{\theta}(s_i\mid s_{<i}),
\end{equation}
and the same factorization generates samples sequentially according to
\begin{align}
    s_1&\sim p_{\theta}(s_1),\qquad
    s_2\sim p_{\theta}(s_2\mid s_1),
    \nonumber\\
    &\hspace{2.5em}\ldots,\qquad
    s_V\sim p_{\theta}(s_V\mid s_{<V}).
\end{align}
For a target Boltzmann distribution
\begin{equation}
    \pi(\boldsymbol{s})
    =
    Z^{-1}e^{-S(\boldsymbol{s})},
\end{equation}
the VAN is trained by minimizing the reverse Kullback--Leibler divergence. Since the normalization constant $Z$ is independent of $\theta$, the corresponding variational objective can be written as
\begin{equation}
    \mathcal{L}_{\mathrm{VAN}}
    =
    \mathbb{E}_{\boldsymbol{s}\sim p_{\theta}}
    \left[
        \log p_{\theta}(\boldsymbol{s})
        +
        S(\boldsymbol{s})
    \right].
    \label{eq:van_objective}
\end{equation}
Equation~\eqref{eq:van_objective} is 
\begin{equation}
     \mathcal{L}_{\mathrm{VAN}} = D_{\rm KL}(p_\theta\Vert\pi)-\log Z,
\end{equation}
 so its minimizer is independent of the unknown normalization $Z$.

Because the site conditionals are normalized, $p_\theta(\boldsymbol{s})$ can be evaluated directly in a Hastings ratio. We hold $\theta$ fixed during Markov-chain sampling, and the acceptance rule below leaves $\pi$ invariant. For a learned single-site update, a site $x$ is selected and a new value is drawn from the corresponding VAN conditional,
\begin{equation}
    s_x'
    \sim
    p_{\theta,x}(\cdot\mid\boldsymbol{s}),
\end{equation}
while all other lattice variables are left unchanged. The proposal is then accepted with probability
\begin{equation}
    \alpha_{\mathrm{AR}\text{-}1}
    (\boldsymbol{s}\rightarrow\boldsymbol{s}')
    =
    \min\left[
        1,
        \frac{
            \pi(\boldsymbol{s}')
            p_{\theta,x}(s_x\mid\boldsymbol{s}')
        }{
            \pi(\boldsymbol{s})
            p_{\theta,x}(s_x'\mid\boldsymbol{s})
        }
    \right].
    \label{eq:van_single_site_mh}
\end{equation}
To vary the spatial scope of the learned proposal, we extend the single-site transition to an ordered block whose sites are regenerated sequentially and accepted or rejected together.

\section{Autoregressive Block Update}
\label{sec:block_update}

We decompose the configuration as $\boldsymbol{s}=(\boldsymbol{s}_B,\boldsymbol{s}_{\bar B})$, where $B=(i_1,\ldots,i_m)$ specifies an ordered block of lattice sites and $\bar B$ denotes its complement, containing all lattice sites outside the block. For AR-$b$ with $1\leq b<L$, its anchor is drawn uniformly at each attempt. Starting from that anchor, the sites of the $b\times b$ block are listed in row-major order relative to the anchor, with both coordinates reduced modulo $L$. Variables outside this ordered list remain unchanged.

Starting from $\boldsymbol{z}^{(0)}=\boldsymbol{s}$, substep $k$ draws $s'_{i_k}$ from the VAN site density evaluated on $\boldsymbol{z}^{(k-1)}$ and replaces site $i_k$ to form $\boldsymbol{z}^{(k)}$. The forward proposal density is 
\begin{equation}
Q_B(\boldsymbol{s}'\mid\boldsymbol{s})
=
\prod_{k=1}^{m}
p_{\theta,i_k}\!\left(
s'_{i_k}\mid\boldsymbol{z}^{(k-1)}
\right),
\qquad
\boldsymbol{s}'=\boldsymbol{z}^{(m)},
\label{eq:block_forward_product}
\end{equation}
with $\boldsymbol{s}'_{\bar B}=\boldsymbol{s}_{\bar B}$.

The reverse product starts from $\widetilde{\boldsymbol{z}}^{(0)} =\boldsymbol{s}'$ and restores the old values in the same site order. If $\widetilde{\boldsymbol{z}}^{(k)}$ is obtained by replacing site $i_k$ with $s_{i_k}$, then
\begin{equation}
Q_B(\boldsymbol{s}\mid\boldsymbol{s}')
=
\prod_{k=1}^{m}
p_{\theta,i_k}\!\left(
s_{i_k}\mid\widetilde{\boldsymbol{z}}^{(k-1)}
\right).
\label{eq:block_reverse_product}
\end{equation}

\begin{figure*}[t]
\centering
\subfloat[Gaussian noise $\epsilon$]{%
\includegraphics[width=0.235\textwidth]{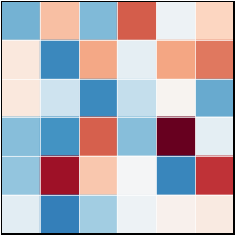}}
\hfill
\subfloat[$12/36$ generated]{%
\includegraphics[width=0.235\textwidth]{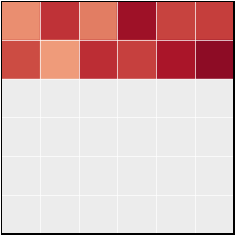}}
\hfill
\subfloat[$24/36$ generated]{%
\includegraphics[width=0.235\textwidth]{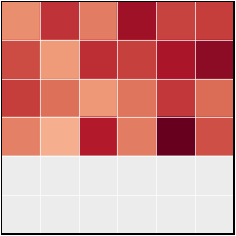}}
\hfill
\subfloat[Direct sample $\boldsymbol{s}$]{%
\includegraphics[width=0.235\textwidth]{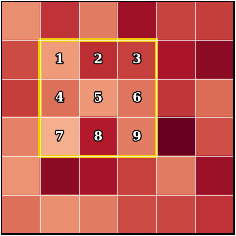}}
\par\vspace{0.6ex}
\makebox[\textwidth][c]{%
\subfloat[$\boldsymbol{z}^{(3)}$]{%
\includegraphics[width=0.235\textwidth]{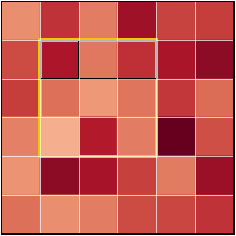}}
\hspace{0.02\textwidth}
\subfloat[$\boldsymbol{z}^{(6)}$]{%
\includegraphics[width=0.235\textwidth]{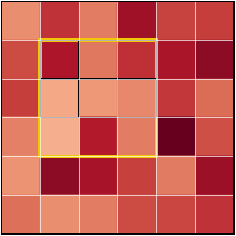}}
\hspace{0.02\textwidth}
\subfloat[Candidate $\boldsymbol{s}'$]{%
\includegraphics[width=0.235\textwidth]{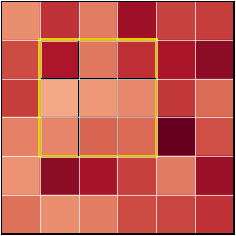}}}
\par\vspace{0.2ex}
\includegraphics[width=0.40\textwidth]{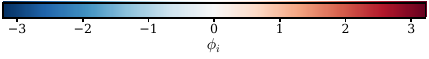}
\caption{Direct VAN generation and one $3\times3$ block update using an
$L=6$ network trained at $\kappa_{\rm train}=0.30$. Panels (a)--(d) show a raster-ordered draw
from $p_\theta$; gray sites have not yet been sampled. The completed draw
in panel (d) is the initial state. Panels (e)--(g) replace the numbered
block sites in order while leaving the remaining field unchanged. The
Hastings ratio uses the forward probabilities accumulated along these
partial configurations and the reverse probabilities obtained by
restoring the original block values in the same order. The shared color
scale applies to panels (b)--(g).}
\label{fig:block_paths}
\end{figure*}

The candidate is accepted with probability
\begin{equation}
\alpha_B(\boldsymbol{s}\to\boldsymbol{s}')
=
\min\!\left[
1,\,
\frac{
\pi(\boldsymbol{s}')Q_B(\boldsymbol{s}\mid\boldsymbol{s}')
}{
\pi(\boldsymbol{s})Q_B(\boldsymbol{s}'\mid\boldsymbol{s})
}
\right].
\label{eq:block_mh_accept}
\end{equation}
For $\pi(\boldsymbol{s})\propto e^{-S(\boldsymbol{s})}$, Eq.~\eqref{eq:block_mh_accept} is evaluated from $S(\boldsymbol{s}')-S(\boldsymbol{s})$ and the two sequential proposal products above.  

Figure~\ref{fig:block_paths} illustrates the construction for a $3\times3$ block.  Starting from the initial configuration, the selected block sites are regenerated sequentially in the prescribed order.  At each substep, the newly generated value is inserted into the current partial configuration and therefore enters the conditional distributions used for all subsequent sites.  The forward proposal density is thus obtained by accumulating the corresponding sitewise conditional probabilities along this sequence. The reverse proposal starts instead from the completed candidate configuration and restores the original block values one by one in the same site order.  Each reverse factor is evaluated before the corresponding old value is restored.  Consequently, the forward generation and reverse restoration generally traverse different partial configurations, and their two sequential probability products must both be retained in the Metropolis--Hastings ratio of Eq.~\eqref{eq:block_mh_accept}.

For $1\leq b<L$, one recorded sweep contains
\begin{equation}
n_b=\left\lceil\frac{L^2}{b^2}\right\rceil
\label{eq:block_attempts_per_sweep}
\end{equation}
independently anchored finite-block attempts. The $b=1$ case is the learned single-site kernel AR-$1$. The global autoregressive independence-MH kernel (global AR-iMH) is defined separately: it regenerates all $L^2$ sites once in the fixed global row-major order, so that
\begin{equation}
Q_{\rm glob}(\boldsymbol{s}'\mid\boldsymbol{s})
=p_\theta(\boldsymbol{s}').
\label{eq:global_proposal}
\end{equation}
Appendix~\ref{app:block_mh_proof} proves normalization and detailed balance for the finite-block construction and its randomized sweep.

For a block that is not a suffix of the global autoregressive order, variables outside the block that occur later in that order do not enter the site conditionals used in the proposal, although their full-joint density factors generally change when the earlier block variables are replaced. Even if $p_\theta(\boldsymbol{s})=\pi(\boldsymbol{s})$ for the full field, this partial update need not be rejection-free; its Hastings ratio still requires both sequential products.

\section{Two-dimensional $\phi^4$ scalar field theory}
\label{sec:phi4}
We study the two-dimensional $\phi^4$ scalar field theory on an $L\times L$ square lattice with periodic boundary conditions. Its Euclidean action is
\begin{equation}
S[\phi] = \sum_{x} \left[
-2\kappa \sum_{\mu=1,2} \phi_x \phi_{x+\hat{\mu}}
+ (1-2\lambda)\phi_x^2 + \lambda\phi_x^4
\right],
\label{eq:phi4_action}
\end{equation}
where $\phi_x\in\mathbb{R}$, $\kappa$ is the hopping parameter, and $\lambda$ is the quartic coupling. We fix $\lambda=0.022$ and denote the lattice volume by $V=L^2$. Earlier lattice studies characterized the phase transition and critical scaling of the two-dimensional $\phi^4$ theory~\cite{Loinaz1998,De2005}. For the hopping-parameter action in Eq.~\eqref{eq:phi4_action}, a susceptibility scan at $\lambda=0.02$ locates the transition near $\kappa\simeq0.27$~\cite{Pawlowski2017}.  At $\lambda=0.022$, subsequent lattice studies place the finite-volume transition region near $\kappa\simeq0.27$--$0.275$~\cite{Pawlowski2020,Nicoli2021,Nicoli2023ModeCollapse}. We therefore use $\kappa=0.273$ for the near-transition finite-volume comparison.

We consider the following observables to characterize the sampled lattice configurations:

Absolute magnetization.
The magnetization per lattice site is defined as
\begin{equation}
M
=
\frac{1}{V}\sum_x \phi_x ,
\end{equation}
and we use
\begin{equation}
\langle |M| \rangle
\end{equation}
as the corresponding observable, avoiding cancellation between the two magnetization sectors related by the global $\mathbb{Z}_2$ symmetry.

Action density.
The action density is defined as
\begin{equation}
    \left\langle \frac{S}{V} \right\rangle .
\end{equation}

Binder cumulant.
The Binder cumulant is defined by
\begin{equation}
    U_4
    =
    1-
    \frac{\langle M^4\rangle}
    {3\langle M^2\rangle^2},
    \label{eq:binder_cumulant}
\end{equation}
and provides a dimensionless measure of the shape of the magnetization distribution.

Finite-volume susceptibility.
The susceptibility associated with the absolute magnetization is defined as
\begin{equation}
    \chi_{|M|}
    =
    V\left(
    \langle M^2\rangle
    -
    \langle |M|\rangle^2
    \right),
    \label{eq:chi_abs}
\end{equation}
where $V$ denotes the total number of lattice sites.

\subsection{Numerical setup}
\label{sec:numerical_setup}
We denote the learned single-site Metropolis--Hastings update by AR-$1$, while AR-$b$ denotes a sequential $b\times b$ block update. A complete AR-$b$ sweep contains the $n_b$ block proposals defined in Eq.~\eqref{eq:block_attempts_per_sweep}. The number of proposed site values in one sweep is therefore $n_b b^2$.

In the $L=6$ coupling scan of Figure~\ref{fig:l6_kappa_summary}, a separate network is trained at each $\kappa=0.20,0.21,\ldots,0.34$, with the target action evaluated at the same coupling.

The main comparison uses AR-$1$, finite sizes drawn from $b\in\{2,3,4,6\}$ whenever $b<L$, and the separately defined global AR-iMH proposal at $L=6$, $8$, and $10$. Thus no finite AR-$6$ kernel is reported at $L=6$.A tuned random-walk single-site Metropolis kernel (RW-$1$) provides a
local comparison, and HMC supplies reference chains for the
corresponding target actions.

Unless stated otherwise, each comparison uses one Markov chain without thinning initialized from a saved HMC configuration. We discard $1{,}000$ burn-in updates and retain $10{,}000$ states. One recorded RW-$1$ sweep contains $L^2$ single-site attempts; one finite AR-$b$ sweep contains the $n_b$ block attempts in Eq.~\eqref{eq:block_attempts_per_sweep}; one global AR-iMH update is a full-field proposal; and one HMC update is a molecular-dynamics trajectory. The HMC runs use trajectory length $\tau_{\rm MD}=1$, $n_{\rm leap}=10$ leapfrog steps, and step size $\varepsilon=0.1$.The production sequence is recorded after every sweep without
discarding states according to their acceptance decisions. If a
complete sweep leaves the configuration unchanged, the repeated state
remains in the retained sequence.

\subsection{Physical observables of the $\phi^4$ theory}
\label{sec:phi4_observables}
For each retained Markov chain, the primary series
$|M|$, $M^2$, $M^4$, and $S/V$ are analyzed separately with
the $\Gamma$-method. We use $S=2.0$ for the automatic summation
window. The action-density uncertainty therefore uses the
autocorrelation of $S/V$ itself. The derived observables $U_4$ and
$\chi_{|M|}$ are evaluated as functions of the correlated estimates of
the primary means. Their uncertainties are obtained using the
$\Gamma$-method, including both autocovariances and cross-covariances
of the relevant primary time series.

For independent VAN samples, we also construct importance-reweighted
estimates. For samples $\boldsymbol{s}_n\sim p_\theta$, the
unnormalized importance weights are
\begin{equation}
w_n=
\frac{e^{-S(\boldsymbol{s}_n)}}
     {p_\theta(\boldsymbol{s}_n)}.
\label{eq:is_weights}
\end{equation}
The corresponding self-normalized importance-sampling estimator is
\begin{equation}
\widehat{\langle O\rangle}_{\rm IS}
=
\frac{\sum_{n=1}^{N}w_n O(\boldsymbol{s}_n)}
     {\sum_{n=1}^{N}w_n}.
\label{eq:snis_estimator}
\end{equation}
For independent direct-VAN samples, uncertainties are propagated from
the joint covariance of the primary sample means using the delta method.
For VAN+IS, we use the self-normalized weighted covariance of the
primary observables. Its centering by the weighted estimates accounts
for fluctuations of the normalization denominator, and the resulting
covariance is propagated to $U_4$ and $\chi_{|M|}$ using the delta method.
The uncertainty of the normalized importance-sampling effective sample
size is estimated from $200$ nonparametric bootstrap resamples.

\begin{figure*}[t]
\centering
\includegraphics[width=0.72\textwidth]{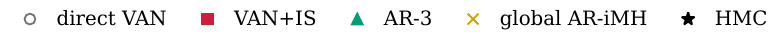}
\par\vspace{-0.2ex}
\subfloat[$\langle|M|\rangle$]{%
\includegraphics[width=0.245\textwidth]{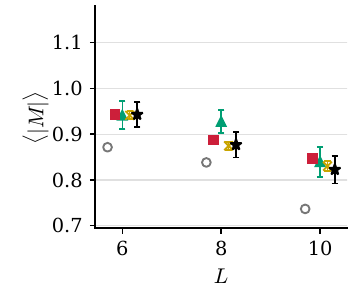}}
\hfill
\subfloat[$\langle S_{\kappa_{\rm target}}/V\rangle$]{%
\includegraphics[width=0.245\textwidth]{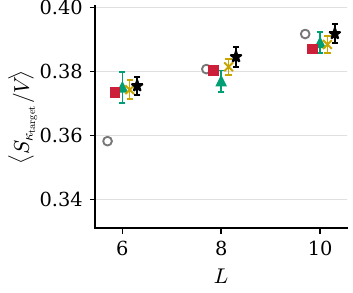}}
\hfill
\subfloat[$U_4$]{%
\includegraphics[width=0.245\textwidth]{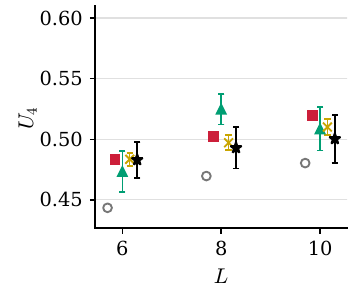}}
\hfill
\subfloat[$\chi_{|M|}$]{%
\includegraphics[width=0.245\textwidth]{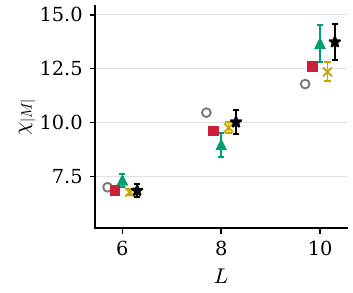}}
\par\vspace{0.7ex}
\subfloat[$\langle|M|\rangle$]{%
\includegraphics[width=0.245\textwidth]{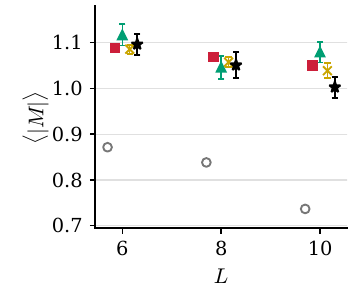}}
\hfill
\subfloat[$\langle S_{\kappa_{\rm target}}/V\rangle$]{%
\includegraphics[width=0.245\textwidth]{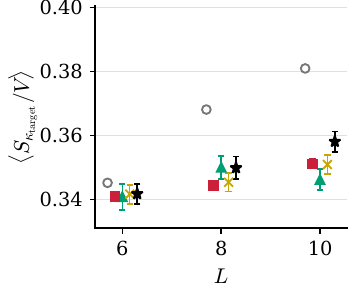}}
\hfill
\subfloat[$U_4$]{%
\includegraphics[width=0.245\textwidth]{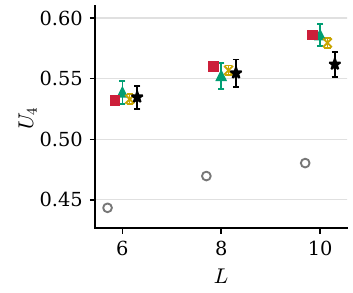}}
\hfill
\subfloat[$\chi_{|M|}$]{%
\includegraphics[width=0.245\textwidth]{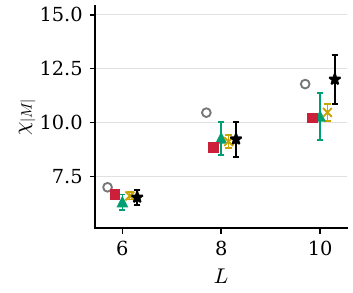}}
\caption{Finite-volume comparison of $\langle|M|\rangle$, $\langle S_{\kappa_{\rm target}}/V\rangle$, $U_4$, and $\chi_{|M|}$. 
Markers denote the sampling method: circles for direct VAN, squares for VAN+IS, upward triangles for AR-$3$, downward triangles for global AR-iMH, and diamonds for HMC. 
The upper and lower rows correspond to $\kappa_{\rm target}=0.270$ and $0.273$, respectively, with all networks trained at $\kappa_{\rm train}=0.270$. Direct VAN and VAN+IS use the same $50{,}000$ independent proposals at each $L$, whereas AR-$3$, global AR-iMH, and HMC are obtained from separate chains. Error bars show one standard error, with uncertainties estimated using the procedures described in the text.}
\label{fig:phi4_observables_vs_L}
\end{figure*}

Figure~\ref{fig:phi4_observables_vs_L} compares the absolute magnetization, action density, Binder cumulant, and finite-volume susceptibility at two target couplings, $\kappa=0.270$ and $0.273$, using networks trained at $\kappa_{\rm train}=0.270$. Direct VAN samples characterize the learned proposal distribution, whereas VAN+IS and the Metropolis--Hastings-corrected autoregressive chains provide corrected estimates for comparison with the HMC reference.

As shown in Figure~\ref{fig:phi4_observables_vs_L}, the corrected estimates remain close to the HMC reference across the three lattice sizes, whereas the direct VAN samples show increasingly visible deviations as $L$ increases. The shift from $\kappa_{\rm target}=0.270$ to $0.273$ is reflected in larger $\langle |M| \rangle$ and $U_4$. Since $\kappa_{\rm target}$ is not provided as an input to the VAN, changing the target coupling leaves the learned proposal distribution unchanged. Consequently, the direct-sample values of $\langle |M| \rangle$, $U_4$, and $\chi_{|M|}$ are identical for the two targets, while the action density, importance weights, and Metropolis--Hastings acceptance probability change through their dependence on the target action. Appendix~\ref{app:target_offset_control} reports the corresponding $L=8$ numerical values, together with the global-proposal acceptance and autocorrelation. The complete kernel-by-kernel values at $\kappa_{\rm target}=0.273$ are reported in Tables~\ref{tab:phi4_observables_allL} and \ref{tab:phi4_susceptibility_allL} of Appendix~\ref{app:complete_observables}.

Figure~\ref{fig:l6_kappa_summary} shows the $L=6$ scan in which the network and target use the same coupling at each point.  Panels~(a)--(d) display the coupling dependence of the four physical observables.  As $\kappa$ increases, $\langle |M| \rangle$ remains small at weak coupling and then rises rapidly through the transition region, signaling the development of an ordered phase.  Correspondingly, the action density $\langle S/V\rangle$ decreases smoothly and becomes increasingly negative at larger $\kappa$.  The Binder cumulant $U_4$ increases from values close to zero in the disordered regime toward its large-$\kappa$ plateau, reflecting the progressive concentration of the magnetization distribution in the ordered phase.  The susceptibility $\chi_{|M|}$ exhibits a pronounced finite-volume peak near the transition region, where fluctuations of the magnetization are maximal.  Across the scan, the Metropolis--Hastings-corrected AR-3 and global AR-iMH estimates, as well as the importance-reweighted VAN+IS results, closely follow the HMC reference for all four observables.
\begin{figure*}[t]
\centering
\includegraphics[width=0.70\textwidth]{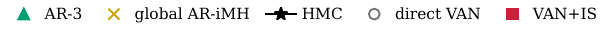}
\par\vspace{-0.2ex}
\subfloat[$\langle|M|\rangle$]{%
\includegraphics[width=0.31\textwidth]{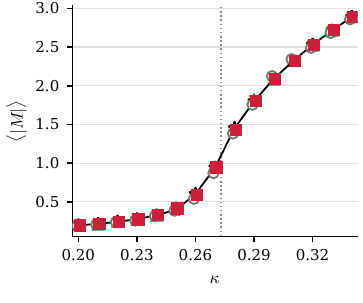}}
\hfill
\subfloat[$\langle S/V\rangle$]{%
\includegraphics[width=0.31\textwidth]{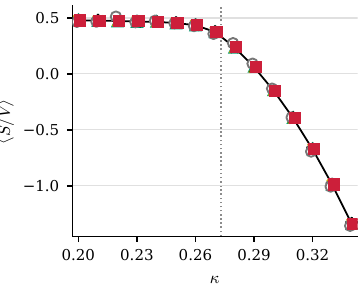}}
\hfill
\subfloat[$U_4$]{%
\includegraphics[width=0.31\textwidth]{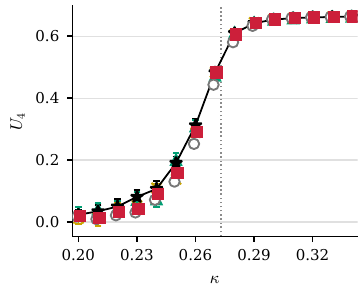}}
\par\vspace{0.8ex}
\subfloat[$\chi_{|M|}$]{%
\includegraphics[width=0.31\textwidth]{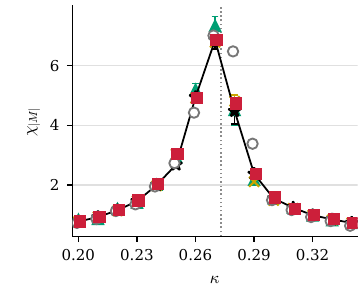}}
\hfill
\subfloat[Acceptance]{%
\includegraphics[width=0.31\textwidth]{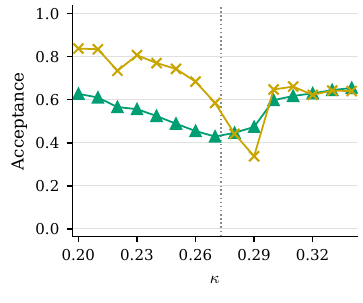}}
\hfill
\subfloat[$N_{\rm eff}^{\rm IS}/N$]{%
\includegraphics[width=0.31\textwidth]{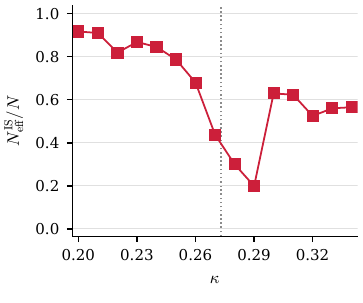}}
\caption{$L=6$ coupling scan at $\lambda=0.022$ for
$\kappa=0.20,0.21,\ldots,0.34$.
Panels (a)--(d) show $\langle|M|\rangle$, $\langle S/V\rangle$, $U_4$, and $\chi_{|M|}$; panel (e) shows the acceptance rate of AR-$3$ and global AR-iMH, and panel (f) shows the normalized importance-sampling effective sample size (ESS) from $50{,}000$ direct VAN proposals. At every point, $\kappa_{\rm train}=\kappa_{\rm target}=\kappa$. AR-$3$ and global AR-iMH are Metropolis--Hastings corrected, VAN+IS is importance reweighted, and HMC supplies the reference in panels (a)--(d). The dotted line marks $\kappa=0.273$, the target coupling of the separate comparison with $\kappa_{\rm train}=0.270$ and $\kappa_{\rm target}=0.273$.}
\label{fig:l6_kappa_summary}
\end{figure*}



All four observables are invariant under $M\mapsto-M$. Table~\ref{tab:sector_transport_L8} therefore also reports the positive-sector fraction and the number of sign changes between consecutive recorded states in the $L=8$ chains.  The quantity $f_+=N(M>0)/N$ measures the relative occupation of the two $\mathbb{Z}_2$-related magnetization sectors, while $N_{\rm cross}$ provides a direct finite-chain measure of how frequently the sampler moves between them.  All of the retained chains visit both signs of the magnetization, but the frequency of inter-sector transitions depends strongly on the update scheme.  In particular, the global AR-iMH chain exhibits substantially more sign changes than the local and finite-block kernels, indicating much more frequent transport between the two symmetry-related sectors over the same number of retained states.  The values of $f_+$ also show that finite chains need not populate the two sectors exactly equally, even when both sectors are repeatedly visited.  These quantities are intended as diagnostics of the observed sector history rather than as equilibrium mixing-time estimators.

\begin{table}[t]
\centering
\begin{tabular}{lrr}
\toprule
Kernel & $f_+=N(M>0)/N$ & $N_{\rm cross}$ \\
\midrule
RW-$1$        & $0.6256$ & $27$ \\
AR-$1$        & $0.4518$ & $73$ \\
AR-$2$        & $0.4492$ & $56$ \\
AR-$3$        & $0.5041$ & $97$ \\
AR-$4$        & $0.4794$ & $60$ \\
AR-$6$        & $0.5996$ & $61$ \\
global AR-iMH & $0.4966$ & $1938$ \\
HMC           & $0.7440$ & $99$ \\
\bottomrule
\end{tabular}
\caption{Signed-magnetization statistics for the $10{,}000$ retained states of each $L=8$ chain at $\kappa_{\rm train}=0.270$ and $\kappa_{\rm target}=0.273$. Here $f_+$ is the fraction with $M>0$, and $N_{\rm cross}$ counts sign changes between consecutive states.}
\label{tab:sector_transport_L8}
\end{table}

\subsection{Block updates versus single-site updates}
\label{sec:block_advantage}

We compare the acceptance rate and integrated autocorrelation time across different autoregressive update sizes, and use the importance-sampling effective sample size to characterize the full-field VAN proposal.

\subsubsection{Acceptance rate}
\label{sec:block_acceptance}

The acceptance rate measures the fraction of proposed updates that are accepted by the Metropolis--Hastings step and therefore provides a diagnostic of the compatibility between the proposal kernel and the target distribution.

Table~\ref{tab:kernel_finite_volume} compares target couplings $0.270$ and $0.273$ at $L=6$, $8$, and $10$, using the network trained at $0.270$ for each volume.

\begin{table*}[t]
\centering
\small
\setlength{\tabcolsep}{3.5pt}
\renewcommand{\arraystretch}{0.98}
\begin{tabular}{lcc@{\qquad}cc@{\qquad}cc}
\toprule
& \multicolumn{2}{c}{$L=6$}
& \multicolumn{2}{c}{$L=8$}
& \multicolumn{2}{c}{$L=10$} \\
\cmidrule(lr){2-3}\cmidrule(lr){4-5}\cmidrule(lr){6-7}
Kernel & Acc. & $\tau_{\rm int}(|M|)$
& Acc. & $\tau_{\rm int}(|M|)$
& Acc. & $\tau_{\rm int}(|M|)$ \\
\midrule
\multicolumn{7}{l}{\textit{(a) $\kappa_{\rm train}=\kappa_{\rm target}=0.270$}} \\
AR-$1$        & $0.800$ & $37.1(9.6)$ & $0.802$ & $32.9(8.1)$  & $0.796$ & $79(26)$ \\
AR-$2$        & $0.601$ & $18.0(3.6)$ & $0.614$ & $29.5(7.0)$  & $0.610$ & $32.4(8.0)$ \\
AR-$3$        & $0.429$ & $22.6(4.9)$ & $0.464$ & $21.3(4.5)$  & $0.463$ & $39.8(10.5)$ \\
AR-$4$        & $0.290$ & $15.3(2.9)$ & $0.325$ & $24.3(5.4)$  & $0.353$ & $35.2(8.9)$ \\
AR-$6$        & -- & --  & $0.155$ & $46.4(12.9)$ & $0.183$ & $33.3(8.3)$ \\
global AR-iMH & $0.585$ & $2.08(17)$  & $0.490$ & $2.62(24)$   & $0.387$ & $5.08(62)$ \\
HMC           & $0.985$ & $18.7(3.8)$ & $0.981$ & $24.7(5.5)$  & $0.978$ & $33.3(8.3)$ \\
\midrule
\multicolumn{7}{l}{\textit{(b) $\kappa_{\rm train}=0.270$, $\kappa_{\rm target}=0.273$}} \\
RW-$1$        & $0.856$ & $193(78)$  & $0.650$ & $75.6(24.3)$ & $0.727$ & $201(82)$ \\
AR-$1$        & $0.800$ & $35.3(9.0)$ & $0.802$ & $56.5(16.7)$ & $0.795$ & $53.3(15.5)$ \\
AR-$2$        & $0.596$ & $19.6(4.0)$ & $0.616$ & $19.9(4.1)$ & $0.603$ & $29.4(7.0)$ \\
AR-$3$        & $0.429$ & $15.3(2.9)$ & $0.457$ & $21.6(4.6)$ & $0.456$ & $25.0(5.6)$ \\
AR-$4$        & $0.278$ & $23.0(5.0)$ & $0.333$ & $25.0(5.6)$ & $0.336$ & $27.0(6.3)$ \\
AR-$6$        & -- & -- & $0.153$ & $62.8(19.1)$ & $0.175$ & $47.1(13.1)$ \\
global AR-iMH & $0.515$ & $2.68(25)$ & $0.396$ & $4.11(46)$ & $0.276$ & $12.7(2.2)$ \\
HMC           & $0.985$ & $14.0(2.5)$ & $0.980$ & $27.6(6.4)$ & $0.977$ & $21.6(4.6)$ \\
\bottomrule
\end{tabular}
\caption{Acceptance and integrated autocorrelation time of $|M|$ for VANs trained at $\kappa_{\rm train}=0.270$. 
Panel (a) uses $\kappa_{\rm target}=0.270$,  and panel (b) uses the same networks with $\kappa_{\rm target}=0.273$. RW-$1$ is shown only in panel (b) because no RW-$1$ run was performed at $\kappa_{\rm target}=0.270$. No finite AR-$6$ run exists at $L=6$; the full-field result is reported only as global AR-iMH. A separate HMC chain is run for each target. The autocorrelation unit is one recorded sweep for RW and autoregressive kernels and one trajectory for HMC; the values are not normalized by computational cost.} 
\label{tab:kernel_finite_volume}
\end{table*}

At $L=8$ and $10$, the acceptance rate decreases systematically as the finite block size increases. For example, at $\kappa_{\rm target}=0.273$, the acceptance at $L=8$ decreases from $0.802$ for AR-$1$ to $0.616$, $0.457$, $0.333$, and $0.153$ for AR-$2$, AR-$3$, AR-$4$, and AR-$6$, respectively. A similar pattern is observed at $L=10$. This behavior reflects the increasing difficulty of jointly proposing a larger number of field variables while maintaining sufficient overlap with the target distribution.

The coupling dependence of the acceptance is further illustrated in panel~(e) of Figure~\ref{fig:l6_kappa_summary} for $L=6$, where $\kappa_{\rm train}=\kappa_{\rm target}$ at every point. Both AR-$3$ and global AR-iMH show a reduction in acceptance as the system approaches the transition region, followed by a recovery at larger $\kappa$. The variation is more pronounced for the global proposal, indicating that full-field updates are more sensitive to changes in the structure of the target distribution across the transition region.

Because the networks are fixed at $\kappa_{\rm train}=0.270$, changing the target to $\kappa_{\rm target}=0.273$ provides a direct test of the sensitivity of each proposal to proposal--target mismatch. The finite-block acceptances change only modestly at fixed $L$ and block size, whereas the global AR-iMH proposal is more strongly affected. Its acceptance decreases from $0.585$ to $0.515$ at $L=6$, from $0.490$ to $0.396$ at $L=8$, and from $0.387$ to $0.276$ at $L=10$. The stronger suppression at larger volume is consistent with the greater sensitivity of a full-field proposal to discrepancies accumulated across the lattice. HMC acceptance remains close to unity for both target couplings.

\subsubsection{Integrated autocorrelation time}
\label{sec:block_tau}

The integrated autocorrelation time quantifies how strongly successive configurations in a Markov chain remain correlated and therefore how rapidly the sampler explores the target distribution. Near a critical point, long-range fluctuations typically lead to an increase in the autocorrelation time, giving rise to critical slowing down.

For the absolute magnetization,  the normalized autocorrelation function is estimated as
\begin{equation}
\rho_{|M|}(t)
=
\frac{\Gamma_{|M|}(t)}
     {\Gamma_{|M|}(0)},
\label{eq:autocorrelation_function}
\end{equation}
where $\Gamma_{|M|}(t)$ is the autocovariance of $|M|$ at lag $t$. The corresponding integrated autocorrelation time is
\begin{equation}
\tau_{\mathrm{int}}(|M|)
=
\frac{1}{2}
+
\sum_{t=1}^{W}\rho_{|M|}(t),
\label{eq:tau_int}
\end{equation}
where the summation window $W$ is determined automatically using the $\Gamma$-method with window parameter $S=2.0$~\cite{Wolff2004,Wolff2007Erratum,Joswig2023}.  The same $\Gamma$-method is used to estimate the statistical uncertainty of the Markov-chain mean of $|M|$.

As shown in Table~\ref{tab:kernel_finite_volume}, the global AR-iMH proposal gives the smallest $\tau_{\rm int}(|M|)$ among the VAN-based kernels at all volumes and target couplings considered. When the networks trained at $\kappa_{\rm train}=0.270$ are applied without retraining to $\kappa_{\rm target}=0.273$, the global AR-iMH autocorrelation time increases from $2.08(17)$ to $2.68(25)$ at $L=6$, from $2.62(24)$ to $4.11(46)$ at $L=8$, and from $5.08(62)$ to $12.7(2.2)$ at $L=10$. The increasingly pronounced deterioration with volume parallels the reduction in global-proposal acceptance, indicating that discrepancies between the learned full-field proposal and the target distribution accumulate more strongly as the lattice size increases. The corresponding autocorrelation functions are shown in Appendix~\ref{app:acf_functions}.

\subsubsection{Importance-sampling effective sample size}
\label{sec:importance_ess}

The effective sample size (ESS) quantifies how many effectively useful samples remain after importance reweighting and therefore provides a diagnostic of the overlap between the full-field VAN proposal and the target distribution. We use the normalized Kish ESS~\cite{Kish1965},
\begin{equation}
\frac{N_{\rm eff}^{\rm IS}}{N}
=
\frac{\left(\sum_{n=1}^{N} w_n\right)^2}
     {N\sum_{n=1}^{N} w_n^2}.
\label{eq:is_ess}
\end{equation}
A value close to unity corresponds to relatively uniform importance weights and strong proposal--target overlap, whereas a small value indicates weight concentration and poor overlap.

For networks trained at $\kappa_{\rm train}=0.270$ and evaluated at $\kappa_{\rm target}=0.273$, the normalized ESS decreases from $0.376(21)$ at $L=6$ to $0.228(13)$ at $L=8$ and $0.0675(94)$ at $L=10$. This systematic decrease indicates that the overlap between the learned full-field proposal and the target distribution deteriorates as the lattice volume increases. The quoted uncertainties are the standard deviations over $200$ nonparametric bootstrap resamples of the $N=50{,}000$ independent VAN proposals. The corresponding $L=6$ coupling scan with $\kappa_{\rm train}=\kappa_{\rm target}$ is shown in Figure~\ref{fig:l6_kappa_summary}(f).

\subsubsection{Block-size dependence}
\label{sec:block_size_dependence}

The dependence on block size reflects a trade-off between the spatial extent of each proposal and its Metropolis acceptance. Increasing the block size allows a single accepted update to modify a larger region of the lattice and can therefore reduce the autocorrelation relative to the learned single-site update. This improvement is not monotonic, however, because larger blocks are also more difficult to propose with sufficient overlap with the target distribution.

This behavior is visible in Table~\ref{tab:kernel_finite_volume}. For both target couplings, the intermediate block sizes AR-$2$--AR-$4$ generally give smaller point estimates of $\tau_{\rm int}(|M|)$ than AR-$1$, whereas increasing the block size further does not produce a systematic additional reduction. In particular, at $L=8$ and $10$, AR-$6$ has substantially lower acceptance while its autocorrelation time is comparable to, or larger than, those of the intermediate-block kernels. The finite-block results therefore indicate a broad intermediate regime in which the gain from updating more sites can compensate for the accompanying loss in acceptance, rather than a monotonic improvement with increasing $b$.

The global AR-iMH update represents the opposite limit, in which the entire field is proposed at once. Within the lattice sizes considered here, it still gives the smallest $\tau_{\rm int}(|M|)$ among the VAN-based kernels. At the shifted target $\kappa_{\rm target}=0.273$, however, the separation between the best finite-block and global autocorrelation estimates becomes smaller as the volume increases, while the full-field importance-sampling ESS simultaneously decreases. This indicates that the advantage of the global proposal becomes increasingly sensitive to the degradation of full-field proposal--target overlap at larger volume.

The present data therefore show no crossover in which a finite-block kernel clearly outperforms the global AR-iMH proposal for $L\leq10$. They do, however, suggest that the competition between reduced full-field overlap and the greater robustness of localized updates may become more important at larger volume or under stronger proposal--target mismatch. Testing this regime would require simulations beyond the lattice sizes and coupling shift considered here.

\section{Discussion and conclusion}
\label{sec:conclusion}

For a block that is not a suffix of the autoregressive order, forward generation and reverse restoration traverse different partial configurations. Evaluating both sequential products gives a reversible fixed-block transition, and independently sampled anchors preserve detailed balance over a finite-block sweep.

At $L=8$, omitting the proposal-density ratio changes $33.1\%$ of paired accept--reject decisions, demonstrating that the sequential forward and reverse proposal probabilities are quantitatively important. Across the finite-block kernels, the results reveal a trade-off between proposal scope and acceptance: intermediate blocks reduce the autocorrelation relative to the learned single-site update, whereas enlarging the block further does not provide a systematic additional gain.

The global AR-iMH proposal remains the most rapidly decorrelating VAN-based kernel in the volumes studied, but it is more sensitive to proposal--target mismatch. Shifting the target from $\kappa=0.270$ to $0.273$ reduces the global acceptance and increases $\tau_{\rm int}(|M|)$ at all three volumes, with the strongest degradation at $L=10$, while the finite-block acceptances change only modestly. The simultaneous decrease of the full-field importance-sampling ESS with volume further indicates a progressive deterioration of full-field proposal--target overlap as the lattice size grows. The present evidence is limited to $L\leq10$ and to autocorrelation measured in each algorithm's native update unit; a comparison of computational efficiency would additionally require wall-clock or operation-count measurements.

\section{Acknowledgments}
The authors thank Gert Aarts, Jianhui Zhang, and Kai Zhou for useful discussions. We also thank the Southern Nuclear Science Computing Center (SNSC) for providing computing resources. SYC is supported by the China Scholarship Council (No. 202308420042) and a Swansea University joint PhD project.
OpenAI Codex (GPT-5, accessed August 2026) assisted with manuscript
revision, LaTeX editing, code review, and consistency checks of numerical
analyses and references. The authors specified each task and checked the
output against the underlying calculations, code, data, and
cited sources, and take responsibility for the final manuscript.

\section{Data availability}
Code, compact run metadata, analysis scripts, and numerical tests will be
released at \url{https://github.com/qxxmax/blockmh}. Trained network
parameters, direct-proposal samples, and unthinned
Markov-chain time series will be deposited in a versioned public data
archive upon publication. During peer review, the complete release package
is available from the corresponding author upon reasonable request while
the public deposits are finalized.

\appendix
\raggedbottom
\section{Network architecture and training}
\label{app:architecture}
\label{app:training}

\paragraph{Autoregressive architecture.}
The $\phi^4$ VAN uses the global row-major raster ordering and has three $7\times7$ masked-convolution layers with 64 channels.  The first layer uses a type-A mask, which excludes the current site and all later sites; the two subsequent layers use type-B masks, which retain the current hidden feature while excluding later sites.  All masked convolutions use zero padding. Periodicity enters the target action and the geometric block coordinates rather than the neural convolution. PReLU activations are used between convolutions. The Gaussian output head produces $(\mu_{ij},\log\sigma_{ij})$ and parameterizes the Gaussian conditional
\begin{equation}
p_\theta(\phi_{ij}\mid\phi_{<ij})
=
\mathcal N(\mu_{ij},\sigma_{ij}^{2}).
\end{equation}
The predicted $\log\sigma_{ij}$ is restricted to the interval $[-10,2]$. The masks make the conditional parameters at each site independent of the current and later raster variables. Direct generation and full-density
evaluation use the same factorization,
\begin{equation}
\log p_\theta(\boldsymbol{\phi})
=
\sum_{ij}\log p_{\theta,ij}(\phi_{ij}\mid\phi_{<ij}).    
\end{equation}

\paragraph{Training.}
The checkpoints used for the reported $L=6$, $8$, and $10$ results were
trained at $\kappa_{\rm train}=0.270$.  The transfer-training lineage was
$L=6\rightarrow8\rightarrow9\rightarrow10$.  The $L=6$ checkpoint was
trained from scratch.  The $L=8$ checkpoint was initialized from $L=6$,
the intermediate $L=9$ checkpoint was initialized from $L=8$, and the
$L=10$ checkpoint was initialized from $L=9$.  The $L=9$ checkpoint was
used only to initialize the reported $L=10$ network; no $L=9$ physical
observable is reported.

Optimization used Adam with $(\beta_1,\beta_2)=(0.9,0.999)$, learning rate
$10^{-3}$, batch size $1000$, no learning-rate schedule, random seed $42$,
and gradient-norm clipping at $1.0$.  At optimization step $t$, the action
term in Eq.~\eqref{eq:van_objective} was multiplied by $1-0.998^t$.
Because the convolutional parameters are shared across lattice sites, the
same weights can initialize a checkpoint at a different $L$.  The step
counts in Table~\ref{tab:van_training_settings} are the optimization steps
performed directly at the listed lattice size rather than cumulative counts
along the transfer-training lineage.

For the $L=6$ coupling scan, a separate network was trained from scratch at each $\kappa=0.20,0.21,\ldots,0.34$ for $5000$ steps, using the same optimizer, annealing schedule, batch size, random seed, and gradient clipping.

\begin{table}[htbp]
  \centering
  \small
  \begin{tabular}{lccc}
    \toprule
    Lattice size & Initialization & $\kappa_{\mathrm{train}}$ & Steps at listed $L$ \\
    \midrule
    $6$  & From scratch      & $0.270$ & $5000$ \\
    $8$  & $L=6$ checkpoint  & $0.270$ & $2500$ \\
    $9$  & $L=8$ checkpoint  & $0.270$ & $2500$ \\
    $10$ & $L=9$ checkpoint  & $0.270$ & $2500$ \\
    \bottomrule
  \end{tabular}
  \caption{Transfer-training lineage and optimization steps at each lattice
  size.  The $L=9$ checkpoint is an intermediate checkpoint used only to
  initialize the reported $L=10$ network; no $L=9$ physical observable is
  reported.}
  \label{tab:van_training_settings}
\end{table}

\paragraph{Random-walk baseline.}
Each RW-$1$ attempt selects a lattice site $x$ uniformly and proposes $\phi_x'=\phi_x+\sigma\xi$, where $\xi\sim\mathcal N(0,1)$; all other field components remain fixed. The symmetric proposal is accepted with probability $\min[1,\exp(-\Delta S)]$, where $\Delta S=S[\phi']-S[\phi]$. One recorded sweep contains $L^2$ such attempts. At $\kappa_{\rm target}=0.273$, $\sigma$ was selected independently at each $L$ from $0.1$, $0.2$, $0.3$, $0.4$, $0.6$, and $0.8$. Each candidate used 250 burn-in and 1000 pilot sweeps. Selection maximized $N/[2\tau_{\rm int}(|M|)t_{\rm wall}]$, with $N=1000$, giving $\sigma=0.3$, $0.8$, and $0.6$ for $L=6$, 8, and 10, respectively.
\paragraph{Recorded-sweep convention.}
For a finite AR-$b$ kernel, one recorded sweep contains
$n_b=\lceil L^2/b^2\rceil$ independently anchored block proposals.
Because the anchor is sampled independently at every attempt, these blocks
do not form a deterministic nonoverlapping tiling of the lattice: blocks
may overlap within a sweep, and a given site may be proposed more than once
or not at all.  Thus $n_b b^2$ is the number of proposed site values in a
recorded sweep, not the number of distinct lattice sites touched.

For RW-$1$, one recorded sweep contains $L^2$ single-site proposals; one
global AR-iMH update is one full-field proposal; and one HMC update is one
molecular-dynamics trajectory.  Autocorrelation times are reported in these
native update units.  Operation counters and hardware-specific wall times
are retained in the run metadata as provenance, but are not used here as
matched-cost efficiency measures.
\section{Detailed balance for the sequential autoregressive block-MH kernel}
\label{app:block_mh_proof}

Fix an ordered block $B=(i_1,\ldots,i_m)$.  Let $\lambda_i$ be Lebesgue measure for the field at site $i$, and define
\[
\lambda_B=\bigotimes_{i\in B}\lambda_i,
\qquad
\eta=\bigotimes_i\lambda_i.
\]
The target measure is $\pi(d\boldsymbol{s}) =\pi(\boldsymbol{s})\eta(d\boldsymbol{s})$. A block proposal changes only the sites in $B$, so every proposed configuration satisfies $\boldsymbol{s}'_{\bar B}=\boldsymbol{s}_{\bar B}$.

\paragraph{Proposal normalization and reverse path.} 
For a fixed configuration $\boldsymbol{s}$, let $\boldsymbol{u}=(\boldsymbol{u}_B,\boldsymbol{s}_{\bar B})$.  The proposal density in Eq.~\eqref{eq:block_forward_product} is
\begin{equation}
Q_B(\boldsymbol{u}\mid\boldsymbol{s})
=
\prod_{k=1}^{m}
p_{\theta,i_k}\!\left(
u_{i_k}\mid
\boldsymbol{z}^{(k-1)}
\right),
\label{eq:block_density_appendix}
\end{equation}
where $\boldsymbol{z}^{(k-1)}$ contains the proposed values at $i_1,\ldots,i_{k-1}$ and the original values at the remaining sites. Because every site conditional is normalized, successive integration over $u_{i_m},\ldots,u_{i_1}$ gives
\begin{equation}
\int_{\mathbb R^m}
Q_B(\boldsymbol{u}\mid\boldsymbol{s})
\,\lambda_B(d\boldsymbol{u}_B)
=1.
\label{eq:block_proposal_normalization}
\end{equation}
The corresponding proposal measure is
\begin{equation}
Q_B(d\boldsymbol{s}'\mid\boldsymbol{s})
=
Q_B(\boldsymbol{s}'\mid\boldsymbol{s})
\,\lambda_B(d\boldsymbol{s}'_B)
\,\delta_{\boldsymbol{s}_{\bar B}}
(d\boldsymbol{s}'_{\bar B}).
\label{eq:block_proposal_kernel}
\end{equation}
Starting from $\boldsymbol{s}'$, the reverse density
$Q_B(\boldsymbol{s}\mid\boldsymbol{s}')$ is evaluated in the same order $i_1,\ldots,i_m$, with each factor evaluated before the corresponding old value is restored.  Reversing the site order would define a different proposal kernel.

\paragraph{Fixed-block detailed balance.}
Let $\alpha_B$ be the acceptance probability in Eq.~\eqref{eq:block_mh_accept}.  Including rejected proposals, the transition kernel is

\begin{align}
K_B(d\boldsymbol{s}'\mid\boldsymbol{s})
={}&
Q_B(d\boldsymbol{s}'\mid\boldsymbol{s})
\alpha_B(\boldsymbol{s},\boldsymbol{s}')
+\rho_B(\boldsymbol{s})
\delta_{\boldsymbol{s}}(d\boldsymbol{s}'),
\nonumber\\
\rho_B(\boldsymbol{s})
={}&
1-\int
Q_B(d\boldsymbol{u}\mid\boldsymbol{s})
\alpha_B(\boldsymbol{s},\boldsymbol{u}).
\label{eq:block_complete_kernel}
\end{align}
For distinct configurations connected by the proposal,
\begin{align}
&\pi(\boldsymbol{s})
Q_B(\boldsymbol{s}'\mid\boldsymbol{s})
\alpha_B(\boldsymbol{s},\boldsymbol{s}')
\nonumber\\
&\qquad=
\min\!\left\{
\pi(\boldsymbol{s})
Q_B(\boldsymbol{s}'\mid\boldsymbol{s}),
\pi(\boldsymbol{s}')
Q_B(\boldsymbol{s}\mid\boldsymbol{s}')
\right\}
\nonumber\\
&\qquad=\pi(\boldsymbol{s}')
Q_B(\boldsymbol{s}\mid\boldsymbol{s}')
\alpha_B(\boldsymbol{s}',\boldsymbol{s}).
\label{eq:block_detailed_balance_appendix}
\end{align}
Equation~\eqref{eq:block_detailed_balance_appendix} shows that the $\pi$-weighted accepted proposal flow is symmetric under $\boldsymbol{s}\leftrightarrow\boldsymbol{s}'$. The $\pi$-weighted rejection measure is supported on $\boldsymbol{s}'=\boldsymbol{s}$ and is also symmetric. Hence
\begin{equation}
\pi(d\boldsymbol{s})
K_B(d\boldsymbol{s}'\mid\boldsymbol{s})
=
\pi(d\boldsymbol{s}')
K_B(d\boldsymbol{s}\mid\boldsymbol{s}'),
\end{equation}
and integration over either configuration gives $\pi K_B=\pi$.

\paragraph{Randomized blocks and recorded sweeps.}
For $1\leq b<L$, let $\nu_b(B)$ be the state-independent distribution obtained by drawing the periodic block anchor uniformly, and define $K_b(d\boldsymbol{s}'\mid\boldsymbol{s}) =\sum_B\nu_b(B)K_B(d\boldsymbol{s}'\mid\boldsymbol{s})$. Fixed-block detailed balance gives
\begin{align}
&\pi(d\boldsymbol{s})
K_b(d\boldsymbol{s}'\mid\boldsymbol{s})
\nonumber\\
&\quad=\sum_B\nu_b(B)\,
\pi(d\boldsymbol{s})
K_B(d\boldsymbol{s}'\mid\boldsymbol{s})
\nonumber\\
&\quad=\pi(d\boldsymbol{s}')
K_b(d\boldsymbol{s}\mid\boldsymbol{s}').
\label{eq:block_mixture_detailed_balance}
\end{align}
Each recorded sweep draws a fresh block independently from $\nu_b$ for $n_b=\lceil L^2/b^2\rceil$ attempts, so its kernel is $K_b^{n_b}$. Every power of a reversible kernel is reversible; hence the recorded-sweep kernel satisfies detailed balance with respect to $\pi$.

\paragraph{Global proposal.}
Global AR-iMH is defined separately by generating the complete field once in the fixed global row-major order. Its proposal density is $Q_{\rm glob}(\boldsymbol{s}'\mid\boldsymbol{s}) =p_\theta(\boldsymbol{s}')$, as in Eq.~\eqref{eq:global_proposal}. Substitution into Eq.~\eqref{eq:block_mh_accept} gives the usual independence-MH rule, and the minimum identity in Eq.~\eqref{eq:block_detailed_balance_appendix} proves detailed balance. Section~\ref{sec:numerical_validation} compares the correct AR-$3$ acceptance rule with a rule that omits the proposal-density ratio on the same proposed transitions.

Because the Gaussian heads used here have finite, nonzero variance, their sitewise densities are strictly positive on $\mathbb R$. Since every site belongs to a block selected with positive probability, repeated randomized block updates are $\eta$-irreducible; the global autoregressive proposal has the same full-support property.

\subsection{Effect of the Hastings proposal ratio}
\label{sec:numerical_validation}

At $L=8$, we generate AR-$3$ proposals using the network trained at $\kappa_{\rm train}=0.270$ and evaluate them for $\kappa_{\rm target}=0.273$. For each proposed transition and a shared uniform random variate, we compare the full Hastings rule with a diagnostic rule that omits the proposal-density ratio. The diagnostic rule is evaluated only for this paired comparison; state updates use the full rule.

Omitting $Q_B(\boldsymbol{s}\mid\boldsymbol{s}')/ Q_B(\boldsymbol{s}'\mid\boldsymbol{s})$ changes $33.1\%$ of the paired accept--reject decisions. The log proposal-density difference has standard deviation $3.15$, with central $90\%$ interval $[-5.74,4.57]$. The corrected chain gives $\langle|M|\rangle=1.046(25)$, consistent with the HMC reference $1.051(28)$.

\begin{center}
\begin{minipage}{0.96\columnwidth}
\centering
\refstepcounter{table}
\label{tab:proposal_ratio_ablation}
\begin{tabular}{lcc}
\toprule
Decision rule & Acceptance & Changed decisions \\
\midrule
Full block-MH rule & $0.4567$ & $0$ \\
Without $Q_{\rm rev}/Q_{\rm fwd}$ & $0.4953$ & $0.3315$ \\
\bottomrule
\end{tabular}
\par\vspace{4pt}
\raggedright\small TABLE~\Roman{table}. Paired comparison of the two acceptance rules for AR-$3$ at $L=8$, using a checkpoint trained at $\kappa_{\rm train}=0.270$ and the target at $\kappa_{\rm target}=0.273$. Both rules are evaluated on the same proposed transitions and uniform random numbers. The final column is the fraction
of decisions that differ from the full block-MH rule.
\end{minipage}
\end{center}

\onecolumngrid
\section{Dependence on the target coupling}
\label{app:target_offset_control}

For each volume, the same checkpoint trained at $\kappa_{\rm train}=0.270$ is evaluated at $\kappa_{\rm target}=0.270$ and $0.273$, without retraining. Table~\ref{tab:kernel_finite_volume} gives the corresponding acceptance rates and autocorrelation estimates for the sampling kernels. Figure~\ref{fig:phi4_observables_vs_L} compares the equilibrium observables at both targets, and Table~\ref{tab:l8_target_coupling_closure} reports the $L=8$ values.

\begin{center}
\begin{minipage}{\textwidth}
\centering
\refstepcounter{table}
\label{tab:l8_target_coupling_closure}
\begin{minipage}[t]{0.49\textwidth}
\centering
\scriptsize
\textit{(a) $\kappa_{\rm target}=0.270$}\\[0.25em]
\setlength{\tabcolsep}{3.2pt}
\renewcommand{\arraystretch}{0.96}
\begin{tabular}{lcccc}
\toprule
Estimator
& $\langle|M|\rangle$
& $\langle S/V\rangle$
& $U_4$
& $\chi_{|M|}$ \\
\midrule
Direct VAN
& $0.8382(18)$ & $0.38071(42)$ & $0.4696(13)$ & $10.465(52)$ \\
VAN+IS
& $0.8872(29)$ & $0.38036(74)$ & $0.5022(18)$ & $9.609(84)$ \\
AR-$3$ block MH
& $0.928(24)$ & $0.3769(33)$ & $0.525(13)$ & $8.97(56)$ \\
global AR-iMH
& $0.8745(90)$ & $0.3814(24)$ & $0.4971(62)$ & $9.76(26)$ \\
HMC
& $0.877(28)$ & $0.3845(31)$ & $0.493(17)$ & $10.02(56)$ \\
\bottomrule
\end{tabular}
\end{minipage}\hfill
\begin{minipage}[t]{0.49\textwidth}
\centering
\scriptsize
\textit{(b) $\kappa_{\rm target}=0.273$}\\[0.25em]
\setlength{\tabcolsep}{3.2pt}
\renewcommand{\arraystretch}{0.96}
\begin{tabular}{lcccc}
\toprule
Estimator
& $\langle|M|\rangle$
& $\langle S/V\rangle$
& $U_4$
& $\chi_{|M|}$ \\
\midrule
Direct VAN
& $0.8382(18)$ & $0.36807(43)$ & $0.4696(13)$ & $10.465(52)$ \\
VAN+IS
& $1.0690(33)$ & $0.34430(88)$ & $0.5599(13)$ & $8.845(93)$ \\
AR-$3$ block MH
& $1.046(25)$ & $0.3500(35)$ & $0.552(11)$ & $9.28(77)$ \\
global AR-iMH
& $1.057(11)$ & $0.3453(29)$ & $0.5568(40)$ & $9.12(30)$ \\
HMC
& $1.051(28)$ & $0.3498(36)$ & $0.555(11)$ & $9.23(80)$ \\
\bottomrule
\end{tabular}
\end{minipage}
\vspace{0.45em}
\begin{minipage}{0.90\textwidth}
\small
\textbf{TABLE~\thetable.} The four physical observables at $L=8$ for one checkpoint trained at $\kappa_{\rm train}=0.270$ and two target actions. Direct VAN uses the same $50{,}000$ unweighted proposals in both panels; the action and VAN+IS weights are evaluated for the indicated target. The autoregressive and HMC chains sample that target.
\end{minipage}
\end{minipage}
\end{center}

\section{Autocorrelation functions}
\label{app:acf_functions}

\begin{figure}[!htb]
\centering
\includegraphics[width=0.60\textwidth]{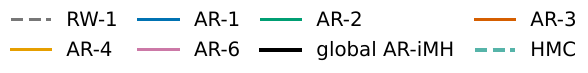}\\[-0.3em]
\subfloat[$L=6$]{%
\label{fig:acf_k270_L6}
\includegraphics[width=0.325\textwidth]{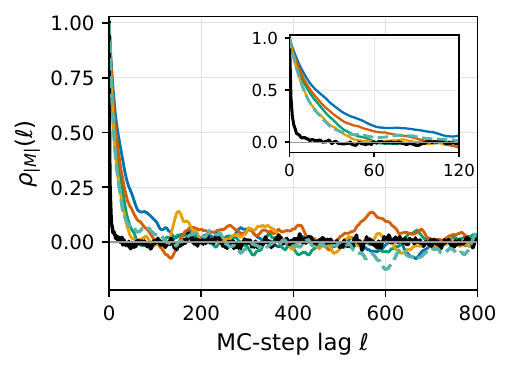}}
\hfill
\subfloat[$L=8$]{%
\label{fig:acf_k270_L8}
\includegraphics[width=0.325\textwidth]{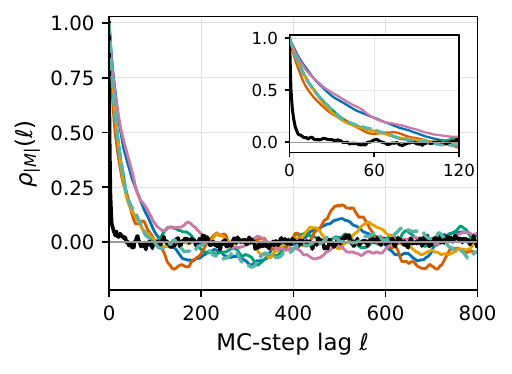}}
\hfill
\subfloat[$L=10$]{%
\label{fig:acf_k270_L10}
\includegraphics[width=0.325\textwidth]{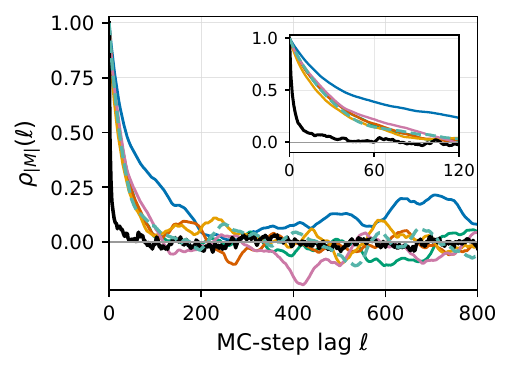}}\\[-0.4em]
\subfloat[$L=6$]{%
\label{fig:acf_k273_L6}
\includegraphics[width=0.325\textwidth]{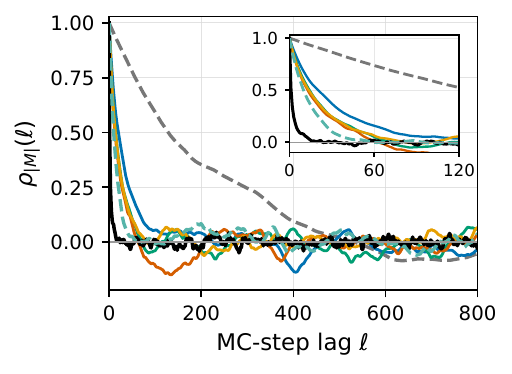}}
\hfill
\subfloat[$L=8$]{%
\label{fig:acf_k273_L8}
\includegraphics[width=0.325\textwidth]{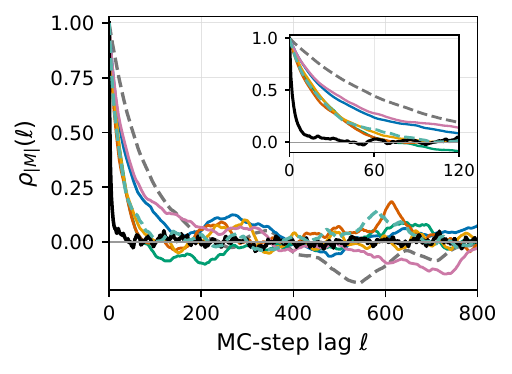}}
\hfill
\subfloat[$L=10$]{%
\label{fig:acf_k273_L10}
\includegraphics[width=0.325\textwidth]{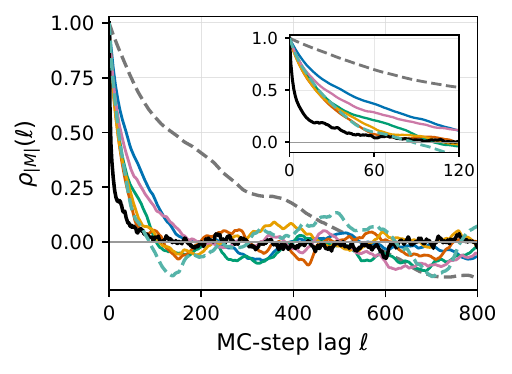}}
\caption{Sample autocorrelation function $\rho_{|M|}(\ell)$ for $L=6$, $8$, and $10$. Panels (a)--(c) use $\kappa_{\rm train}=\kappa_{\rm target}=0.270$; panels (d)--(f) use the same networks with $\kappa_{\rm target}=0.273$. The insets show $\ell\leq120$. No RW-$1$ chain was run at $\kappa_{\rm target}=0.270$.}
\label{fig:acf_target_comparison}
\end{figure}

Figure~\ref{fig:acf_target_comparison} shows the sample autocorrelation functions associated with the integrated estimates in Table~\ref{tab:kernel_finite_volume}. The main axes extend to $\ell=800$, and the insets show $\ell\leq120$. Each curve is computed from one unthinned chain of $10{,}000$ retained states. One lag denotes a finite-block sweep, a global AR-iMH proposal, or an HMC trajectory; for $\kappa_{\rm target}=0.273$, it can also denote an RW-$1$ sweep. Table~\ref{tab:kernel_finite_volume} reports the corresponding $\Gamma$-method estimates. No finite AR-$6$ curve is shown at $L=6$.

\section{Additional observable estimates}
\label{app:complete_observables}

\begin{center}
\refstepcounter{table}
\label{tab:phi4_observables_allL}
\small
\setlength{\tabcolsep}{3.0pt}
\renewcommand{\arraystretch}{1.03}
\resizebox{\textwidth}{!}{%
\begin{tabular}{lccc@{\qquad}ccc@{\qquad}ccc}
\toprule
& \multicolumn{3}{c}{$L=6$}
& \multicolumn{3}{c}{$L=8$}
& \multicolumn{3}{c}{$L=10$}
\\
\cmidrule(lr){2-4}
\cmidrule(lr){5-7}
\cmidrule(lr){8-10}

Kernel
& $\langle|M|\rangle$
& $\langle S/V\rangle$
& $U_4$
& $\langle|M|\rangle$
& $\langle S/V\rangle$
& $U_4$
& $\langle|M|\rangle$
& $\langle S/V\rangle$
& $U_4$
\\
\midrule

RW-$1$
& $1.059(84)$ & $0.3390(96)$ & $0.528(31)$
& $1.064(45)$ & $0.3518(59)$ & $0.560(16)$
& $0.975(75)$ & $0.3591(88)$ & $0.544(36)$
\\

AR-$1$
& $1.081(37)$ & $0.3433(48)$ & $0.529(15)$
& $1.064(41)$ & $0.3473(48)$ & $0.559(17)$
& $1.046(34)$ & $0.3506(42)$ & $0.585(14)$
\\

AR-$2$
& $1.097(27)$ & $0.3415(45)$ & $0.532(11)$
& $1.113(23)$ & $0.3411(37)$ & $0.5725(88)$
& $1.036(25)$ & $0.3512(36)$ & $0.5765(96)$
\\

AR-$3$
& $1.117(23)$ & $0.3408(40)$ & $0.5387(95)$
& $1.046(25)$ & $0.3500(35)$ & $0.552(11)$
& $1.079(23)$ & $0.3462(32)$ & $0.5860(88)$
\\

AR-$4$
& $1.086(30)$ & $0.3465(46)$ & $0.524(13)$
& $1.081(27)$ & $0.3444(43)$ & $0.564(11)$
& $1.062(24)$ & $0.3490(34)$ & $0.5821(86)$
\\

AR-$6$
& -- & -- & --
& $1.020(44)$ & $0.3552(45)$ & $0.540(19)$
& $1.060(30)$ & $0.3494(39)$ & $0.5862(92)$
\\

global AR-iMH
& $1.086(10)$ & $0.3416(30)$ & $0.5330(41)$
& $1.057(11)$ & $0.3453(29)$ & $0.5568(40)$
& $1.039(16)$ & $0.3509(31)$ & $0.5795(39)$
\\

\midrule

HMC
& $1.097(23)$ & $0.3417(31)$ & $0.5346(94)$
& $1.051(28)$ & $0.3498(36)$ & $0.555(11)$
& $1.003(23)$ & $0.3580(32)$ & $0.562(10)$
\\

\bottomrule
\end{tabular}%
}
\vspace{0.5em}
\begin{minipage}{\textwidth}
\small
\textbf{TABLE~\thetable.} Magnetization, action density, and Binder cumulant at $\kappa_{\rm target}=0.273$ for $L=6$, $8$, and $10$, using checkpoints trained at $\kappa_{\rm train}=0.270$. HMC provides the reference values. No finite AR-$6$ run exists at $L=6$; the full-field result is reported in the global AR-iMH row.
\end{minipage}
\end{center}

\vspace{1.2em}
\begin{center}
\refstepcounter{table}
\label{tab:phi4_susceptibility_allL}
\small
\renewcommand{\arraystretch}{1.00}
\begin{tabular}{lccc}
\toprule
Kernel & $\chi_{|M|}$, $L=6$ & $\chi_{|M|}$, $L=8$ & $\chi_{|M|}$, $L=10$ \\
\midrule
RW-$1$        & $6.6(10)$  & $8.5(10)$  & $13.9(27)$ \\
AR-$1$        & $6.94(54)$ & $9.3(13)$  & $10.7(19)$ \\
AR-$2$        & $6.80(43)$ & $8.43(72)$ & $10.3(11)$ \\
AR-$3$        & $6.32(35)$ & $9.28(77)$ & $10.3(11)$ \\
AR-$4$        & $7.19(44)$ & $9.13(90)$ & $10.8(10)$ \\
AR-$6$        & -- & $9.73(89)$ & $9.31(99)$ \\
global AR-iMH & $6.60(16)$ & $9.12(30)$ & $10.47(40)$ \\
\midrule
HMC           & $6.54(36)$ & $9.23(80)$ & $12.0(11)$ \\
\bottomrule
\end{tabular}
\vspace{0.5em}
\begin{minipage}{0.78\textwidth}
\small
\textbf{TABLE~\thetable.} Absolute-magnetization susceptibility $\chi_{|M|}$ at $\kappa_{\rm target}=0.273$ for $L=6$, $8$, and $10$, using checkpoints trained at $\kappa_{\rm train}=0.270$. HMC provides the reference values. No finite AR-$6$ run exists at $L=6$; the full-field result is reported in the global AR-iMH row.
\end{minipage}
\end{center}

\twocolumngrid

\bibliography{ref}

\end{document}